\documentclass[journal]{IEEEtran}

\ifCLASSINFOpdf
\else
   \usepackage[dvips]{graphicx}
\fi
\usepackage{url}

\usepackage{graphicx}
\usepackage{amsmath,empheq}
\usepackage{algorithm}
\usepackage{algpseudocode}
\usepackage{booktabs}
\usepackage{mathtools}
\usepackage[table]{xcolor}
\definecolor{light-gray}{HTML}{FFFFFF}
\definecolor{light-cyan}{HTML}{C4C4C4}

\begin{document}

\title{Efficient Generation of Low Autocorrelation Binary Sequences}

\author{Miroslav Dimitrov, Tsonka Baitcheva \IEEEmembership{Member, IEEE}, and Nikolay Nikolov
\thanks{This work has been partially supported by the Bulgarian National Science Fund under contract number DH 12/8, 15.12.2017.}
\thanks{M. Dimitrov  and T.  Baicheva are with the Institute of Mathematics and Informatics, Bulgarian Academy of Sciences, Sofia, Bulgaria (email:mirdim@math.bas.bg, email:tsonka@math.bas.bg).}
\thanks{N. Nikolov is with the State Agency of National Security, Sofia, Bulgaria.}}

\maketitle

\begin{abstract}
Simple and efficient algorithm based on heuristic search  by shotgun hill climbing to construct binary sequences with small peak sidelobe levels (PSL) is suggested. The algorithm is applied for generation of binary sequences of lengths between 106 and 300. Improvements are obtained in almost half of the considered lengths while for the rest of the lengths, binary sequences with the same PSL values as reported in the state-of-the-art publications are found.
\end{abstract}


\IEEEpeerreviewmaketitle

\section{Introduction}

\IEEEPARstart{S}{equences} with low autocorrelation functions are necessary for a variety of  signal and information processing applications. For example, in pulse codes-based compression for radars and sonars such sequences are used in order to obtain high resolution. The shifts of sequences with low autocorrelation can be also used for better synchronization purposes or to identify users in multi-user systems. Due to their big practical importance, these sequences have been widely studied and various methods for constructing sequences with small values of the autocorrelation are developed. 

The binary sequences of low autocorrelation are of special interest and some of the well known such sequences are the Barker codes \cite{barker1953group}, M-sequences \cite{golomb1967shift}, Gold codes \cite{gold1967optimal}, Kasami codes \cite{kasami1966weight}, Weil sequences \cite{rushanan2006weil}, Legendre sets \cite{pott2006finite} and others (see \cite{levanon2004radar}\cite{skolnik1970radar}). Barker sequences are known to have the best autocorrelation properties, but the longest such sequence is of length 13. M-sequences, Gold codes and Kasami sequences have ideal periodic autocorrelation functions but have no constraints on the sidelobes of their aperiodic autocorrelation functions. As summarized in \cite{nasrabadi2010survey}, during the years a variety of analytical constructions and computer search methods are developed in order to construct binary sequences with relatively small or minimal PSL. By an exhaustive search the minimum values of the PSL for $n\leq 40$\cite{lindner1975binary}, $n\leq 48$\cite{baden1990optimal}, $n=64$\cite{coxson2005efficient}, $n\leq 68$\cite{leukhin2012binary}, $n\leq 74$\cite{leukhin2013optimal},  $n\leq 80$ \cite{leukhin2014exhaustive}, $n\leq 82$ \cite{leukhin2015bernasconi} and $n\leq 84$ \cite{leukhin2017exhaustive}  are obtained. The best currently known values for PSL for $ 85 \leq n \leq 105$ are published in \cite{NC}, and for $n \geq 106$ in \cite{Patent}. In this work we suggest an efficient and easy to implement heuristic algorithm and, as an illustration of its effectiveness, we apply it for generation of binary sequences with lengths between 106 and 300. The generated by us sequences are better, in terms of PSL values, than a significant part of those obtained in \cite{Patent} ones. Our algorithm can also be used for generation of sequences with lengths greater than 300. 

\section{Preliminaries}

Let $B=(b_0,b_1,\cdots ,b_{n-1})$ be a binary sequence of length $n>1$, where $b_i\in \{-1,1\}, 0\leq i\leq n-1$. The \textbf{aperiodic autocorrelation function} (AACF) of $B$ is given by $$ C_u(B)=\sum_{j=0}^{n-u-1} b_jb_{j+u}, \ \ for \ u\in \{0,1,\cdots, n-1\}.$$ We will note that the AACF is originally defined in the interval $\{-n+1,-n+2, \cdots,-2,-1,0,1,2,\cdots,n-1\}$. As the AACF is an even function with $C_u(B)=-C_u(B)$, we will consider it for the interval $\{0,1,\cdots, n-1\}$ only. The $C_0(B)$ is called \textbf{mainlobe} and the rest $C_u(B)$ for $u\in \{1, \cdots ,n-1\}$ are called \textbf{sidelobe} levels. We define the \textbf{peak sidelobe level (PSL)} \cite{turyn1968sequences} of $B$ as $$PSL(B)=\max_{0<u<n} \lvert C_u(B)\rvert.$$

The value of the PSL can also be represented in decibels $$PSL_{db}(B)=20\log \left({\frac{PSL(B)}{n}}\right).$$

Another important measure of an AACF is the \textbf{merit factor} (MF), which gives the ratio of the energy of the mainlobe level to the energy of sidelobe levels, i.e. $$MF(B) = \frac{C_0(B)}{{2\sum_{u=1}^{n-1}\lvert C_u(B)\rvert^2}}.$$

\section{The fitness function choice}
\label{sec:fitness}

Since our goal is to lower the PSL of a given binary sequence, i.e. to lower the value of PSL(B), it makes sense to simultaneously lower the values of each $C_u(B)$, for  $u\in \{1, \cdots ,n-1\}.$ By making this observation, we define the following fitness function: \[ F(B) = \sum_{u=1}^{n-1}{\mathopen| C_u(B) \mathclose| ^P} = \sum_{u=1}^{n-1}{\left(\mathopen| {\sum_{j=0}^{n-u-1} b_jb_{j+u}}\mathclose|\right)^P},\] where $P$ is the magnitude of the fitness function, i.e. the higher the magnitude is the higher the fitness function intolerance to large absolute values of $C_u(B)$'s will be. We made experiments with various values of $P$ and the best results were obtained for values in the interval $\left[ 3, 5 \right]$. Lower values of $P$ makes the fitness function too tolerant to higher absolute values of the PSLs $C_u(B)$, while higher values of $P$ are heavily populating the heuristic topology with local minimums. We have fixed the magnitude P of the fitness function to 4. 

\section{Heuristic search of binary sequences with small PSL by shotgun hill climbing}
\label{sec:SHC}

Let's denote the $i$-th position of a binary sequence $B$ of length $n$ as $b_i$. Flipping the $i$-th position of $B$ is to interchange $b_i$ with $-b_i$. By the neighborhood of the binary sequence $B$, denoted by $N(B)$, we define the set of all binary sequences constructed from $B$ by making a single flip in $B$.

The optimization process takes as input the length of the binary sequence $n$, the fitness function $F$, the threshold value $t$, the two integers $h_{min}$ and $h_{max}$ defining the flipping allowance interval, and the goal $G$ which is the desired final PSL value to be reached.

At the beginning we generate a random binary sequence $B$ of length $n$. Then, by searching the neighborhood of $B$, we look for a better binary sequence, i.e. a binary sequence with smaller fitness value. If some $X$ out of the neighbors of $B$ has PSL equal to $G$ we output $X$ and quit. If during the search of the neighborhood no better binary sequence is found, we are stuck in some local minimum $B'$. In order to escape the local minimum we flip $h$ randomly chosen elements of $B'$, where $h \in \left[h_{min},h_{max}\right]$. We will call such try a \textbf{quake}. In the case when $t$ consecutive quakes are not sufficient to escape the local minimum, we start the process from the beginning by randomly generating new binary sequence, i.e. the shotgun hill climbing approach. The algorithm stops when a binary sequence with the searched value of the PSL is found or when the preliminary defined number of restarts is reached. The pseudo code of the shotgun hill climbing (SHC) algorithm is given in Algorithm \ref{algor:SHC}.

\begin{algorithm}[h]
\caption{SHC algorithm}\label{algor:SHC}
\begin{algorithmic}[1]
\Procedure{HC}{$n,F,t,h_{min},h_{max},G$}
\State $BinSeq \gets R(n)$ \Comment{random binary sequence $BinSeq$ with length $n$}
\State $thresholdLeft \gets t$ \; $bestFit \gets F(BinSeq)$
\State $globFit \gets bestFit$ \; $BinSeqCopy \gets BinSeq$
\Repeat{}	
	\State $NB \gets \Call{$N$}{BinSeq}$ \Comment{generation of all neighbors}
	\State $FLAG \gets True$
	\For{$X \in NB$}
	\If{PSL(X)==G}
		{Output X and Quit}
	\EndIf
	\If{$F(X) < bestFit$} \Comment {a better candidate is found}
		\State $bestFit \gets F(X)$ \; $BinSeq \gets X$ \; $FLAG \gets False$
	\EndIf
	\EndFor
	\If{FLAG}
		\If{$bestFit < globFit$} 
			\State $globFit \gets bestFit$ \; $BinSeqCopy \gets BinSeq$ \; $thresholdLeft \gets t$
		\Else 
			\State {$thresholdLeft \gets thresholdLeft-1$}
			\If{$thresholdLeft > 0$}
				\State {$BinSeq \gets BinSeqCopy$}
				\State {$h \gets \Call{RI}{h_{min},h_{max}$}} \Comment{$h$ is random integer $\in \left[ h_{min},h_{max} \right]$}
				\State \Call{Flip}{$BinSeq,h$} \Comment{flip $h$ random bits in $BinSeq$}
				\State {$bestFit \gets F(BinSeq)$}
			\Else \Comment {the threshold is reached}
				\State $BinSeq \gets R(n)$ \; $thresholdLeft \gets t$ \; $bestFit \gets F(BinSeq)$ \; $globFit \gets bestFit$ \; $BinSeqCopy \gets BinSeq$
			\EndIf
		\EndIf
	
	\EndIf
\Until{STOP condition reached} \Comment{reaching $10^5$ restarts}
\EndProcedure
\end{algorithmic}
\end{algorithm}  

Evidently, the fitness function is the critical resource demanding routine of the algorithm. However, its complexity is comparable to the complexity of the binary sequence PSL calculation itself. The additional negligible overheat is caused by the calculation of the sum of all the $P$-powered mainlobes.

\section{The hyper-parameters choice}
\label{sec:HPC}

The parameter $h_{min}$ should be tolerant to possible optimizations involving any small number of flips. Having this in mind and without any restrictions, we choose $h_{min}=1$. On the other hand, fixing a value of the parameter $h_{max}$ is a trade-off between accuracy and flexibility - smaller values of $h_{max}$ will decrease the algorithm chances to escape from a given local minimum, while higher values of $h_{max}$ will greatly defocus the climbing routines (for example, hoping from hill A to another hill B, before reaching the local minimum of A). During our experiments, we have fixed the value of $h_{max}$ as $\lceil{\sqrt{n}}\;  \rceil$, where $n$ is the length of the starting binary sequence. 

Another important parameter is the threshold value $t$. Choosing a small value of $t$ allows us to restart the process of searching a binary sequence with low PSL value and, instead of loosing more time in trying to escape the current local minimum we have stuck at, we reinitialize the searching procedure by starting from the beginning.

We have tried different meta-heuristic strategies like, for example, simulated annealing method and tabu search. However, it appears that regularly reinitializing the current state of the algorithm, i.e. the core concept of the shotgun hill climbing method, is a more productive strategy to utilize than the aforementioned ones. Evidently, the initial state do matters and by having a low value of $t$ we increase our chances to reinitialize the algorithm from a highly-competitive candidate. During our experiments, we have used a threshold of $t=10^3$.
\section{Results}

We present in Tables \ref{tab:results1}, \ref{tab:results2}, \ref{tab:results3} and \ref{tab:results4} the obtained by Algorithm 1 results for binary sequences of lengths from 106 to 300.  The second column contains the best known by us value of the PSL for the corresponding length. In the third column we present the best value of the PSL obtained by the Algorithm \ref{algor:SHC} and in the fourth the corresponding sequence with this value of the PSL. The sequences are given in a hexadecimal format where -1's are replaced by zeros and the leading -1's are omitted. For example, the binary sequence of length 11 $B=(-1,-1,1,1,-1,1,1,-1,1,1,1)$ is given by $1b7$. The decoding procedure requires the length of the binary sequence. The corresponding values of the PSL in decibels and of the merit factor are calculated and given in the fifth and sixth columns respectively. 

We improve the PSL values for 95 from the included 195 lengths. The remaining 100 binary sequences have the same values of the PSL as the currently known best ones. Furthermore, all of them are unique and unpublished before. 

The suggested in this work algorithm is highly parallelizable so that a multicore architecture can be fully utilized. It is implemented on Python on a single mid-range computer with an octa-core CPU. During our experiments, the time required to reach a given PSL goal was between few minutes to several hours. Furthermore, with each instance of the algorithm, we repeatedly reached binary sequences with lower or same PSL than the state-of-the-art results. 

\begin{table}[h]
\begin{center}
\caption{Results part I}
\label{tab:results1}
\tiny
\ttfamily
\rowcolors{2}{light-gray}{light-cyan}
\begin{tabular}{lccp{3.5cm}cc}
\toprule
$n$ & Old & New & Binary sequence in HEX & db & MF\\
\midrule
\showrowcolors
106 & 7 & 6 & 1366453fff339abc3d613eab4f2 & -24.943 & 5.030\\
107 & 7 & 6 & 3e525b707207bb6280c08c733aa & -25.025 & 4.497\\
108 & 7 & 6 & 9d31b81bc465b48ab7ae0801834 & -25.105 & 5.533\\
109 & 7 & 6 & 1c80e7c337e7ea64d55da750ca5b & -25.186 & 5.636\\
110 & 7 & 6 & 825bebaee519f060d42d81cc8d4 & -23.926 & 5.984\\
111 & 7 & 6 & 1cb387b52c8ed4cfeb048855305c & -24.004 & 5.138\\
112 & 7 & 6 & 68a5614a61368ddf1743207fe706 & -24.082 & 4.931\\
113 & 7 & 6 & 1ae5cb4fe90feae29779ec120644e & -24.160 & 4.409\\
114 & 7 & 6 & 19ed6101bcf959e19a5583a622e81 & -24.236 & 5.375\\
115 & 7 & 7 & 56d413e9ca1c1992f37994f88c502 & -24.312 & 3.952\\
116 & 7 & 7 & 43f475cbd4e3b98d5d0cb6c4840db & -24.387 & 3.925\\
117 & 7 & 7 & 5d8caed643dfa1480b11c347164c1 & -24.462 & 4.108\\
118 & 7 & 7 & 3ce9d9c9ad524fb5f415fade2e1186 & -24.536 & 3.976\\
119 & 7 & 7 & 4b24ce6b455b8b02001de1753c5297 & -24.609 & 4.331\\
120 & 7 & 7 & d91e13e197ad463b9e2d9d5fed2544 & -24.682 & 4.639\\
121 & 7 & 7 & 3fbd241b987f4b8ed966614a888e89 & -24.754 & 4.141\\
122 & 7 & 7 & 28d7ab4e488ce60018781f34d704ae9 & -24.825 & 3.999\\
123 & 8 & 7 & 7d9b6c7bf11e94507c2556d6e6a8c31 & -24.896 & 3.736\\
124 & 8 & 7 & 703ffe14662bdc7cd3f4eb262a49a93 & -24.966 & 4.884\\
125 & 8 & 7 & a8e0e42fc6af59cfb7b640cff64bb2c & -25.036 & 4.134\\
126 & 8 & 7 & ca666b72aa45167f4cc39f00521c2d2 & -25.105 & 4.544\\
127 & 7 & 7 & 73fef5c8d1d95d05cc26917ce097bc2d & -25.174 & 4.815\\
128 & 8 & 7 & 915fca044f23e83a942393ada7bb73e7 & -25.242 & 4.911\\
129 & 8 & 7 & 1856351aa9ada9798eb0070b267d80836 & -25.310 & 3.955\\
130 & 8 & 7 & 7ea03973917046150ca103459afb7b49 & -25.377 & 4.692\\
131 & 8 & 7 & 169633c2e13890d5e540afdd64c811c09 & -25.443 & 4.403\\
132 & 8 & 7 & f4bf06b4afe88af3c79dd76badcd94c8 & -24.350 & 4.431\\
133 & 8 & 7 & 18fe45f33afd90cba4888b9d2b534841e1 & -24.415 & 4.323\\
134 & 8 & 8 & 2b35983c61b4f3bbf3752c69fabe0897a8 & -24.480 & 3.742\\
135 & 8 & 8 & 12c3755bb64459418f4a242e731e1697e & -24.545 & 3.876\\
136 & 8 & 8 & 30ed813f6f583c925aaa2f53e6722f5bcf & -24.609 & 3.925\\
137 & 8 & 8 & d569ca74eebccc573b0208187a6f82fa09 & -24.673 & 4.066\\
138 & 8 & 8 & 3d128917da3431938e6dfd1ef7a2e68bc2f  & -24.736 & 3.771\\
139 & 8 & 8 & 3c0e1d9b35f9bd5342a80db491c406d6f10  & -24.798 & 3.808\\
140 & 8 & 8 & bdcf8e3944f5b152fbbbf01b66a2d0b890a  & -24.861 & 4.026\\
141 & 9 & 8 & 115e1f52e273d156c9af48cc8007b6c649e 5 & -24.923 & 3.923\\
142 & 8 & 8 & 71338901166bd08b7d05ac1a4edf87d1531  & -24.984 & 3.724\\
143 & 8 & 8 & 67aa81c2c56fde794f6365fc0b30db92253 7 & -25.045 & 3.940\\
144 & 8 & 8 & 39716d38490502a3765215eb20ee1bb84ca 3 & -25.105 & 3.886\\
145 & 8 & 8 & 1791bb0ba63bccda7c2a3678dfd6825c792 a0 & -25.166 & 4.477\\
146 & 8 & 8 & 3708999ea4c1f08e12ae8ebcdf092d1215a 20 & -25.225 & 3.975\\
147 & 9 & 8 & 40c48cac0843a2f917ccab14215dd87b792 c7 & -25.285 & 4.122\\
148 & 8 & 8 & 2c24f9cb675dcd540bb0943d629030d83cd c0 & -25.343 & 5.291\\
149 & 8 & 8 & 5a0f857ae7b62266299eee68a141d70085a 58 & -25.402 & 3.698\\
150 & 9 & 8 & 3d63df1b948ddc2689a895072984b2ba7e6 008 & -25.460 & 4.329\\

\bottomrule
\end{tabular}
\end{center}
\end{table}


\begin{table}[h]
\begin{center}
\caption{Results part II}
\label{tab:results2}
\tiny
\ttfamily
\rowcolors{2}{light-gray}{light-cyan}
\begin{tabular}{lccp{3.5cm}cc}
\toprule
$n$ & Old & New & Binary sequence in HEX & db & MF\\
\midrule
\showrowcolors

151 & 8 & 8 & 2640cc90388e31881fe5535d5ac2c2456f2 f16 & -25.518 & 3.999\\
152 & 9 & 8 & 6501a71c13b1fec21d82cfddb2bb3a5d569 536 & -25.575 & 4.068\\
153 & 9 & 8 & 1752e3434eae633cc3817375b05becd5f40 5224 & -25.632 & 3.917\\
154 & 9 & 8 & 3cbe58528eb47f0efe6afbc2ed521dcf988 626d & -25.689 & 3.864\\
155 & 8 & 8 & aaa430985f6a183d3fc9edd8217b0732ef1 b74 & -25.745 & 3.987\\
156 & 9 & 8 & dcaf489c2264f8ff9aeb8d7433f708165a1 6928 & -25.801 & 4.713\\
157 & 8 & 8 & 8c91dbe342975ba661d860071a06d745771 7b0b & -25.856 & 4.241\\
158 & 9 & 8 & 11f07cda85b2c794875eea635521ffdf727 5c666 & -25.911 & 4.479\\
159 & 9 & 8 & 665f717b678d7c472844d61aad3a2e77814 1dbfb & -25.966 & 4.366\\
160 & 9 & 8 & 62088e74b483f5cf4daeb02e3d169de44e9 cd5df & -26.021 & 3.765\\
161 & 9 & 8 & e720b7b8987caaa3ca7e454a0ecc9108245 a5cf & -26.075 & 4.096\\
162 & 9 & 8 & 112db024584a1c7a44aa9b729ab138c0531 f8bf83 & -26.129 & 4.408\\
163 & 9 & 8 & 5af97f061a5a10317fa15510778b32ce219 9c89c2 & -26.182 & 4.104\\
164 & 9 & 8 & 10f81f8297d4226c9428d39b575b9cab2f3 f9a18a & -26.235 & 4.189\\
165 & 9 & 8 & b8089b446cab2ffa99c97939df6953879e4 6bc6f8 & -26.288 & 4.394\\
166 & 9 & 8 & 856dff4fad1b0b93a6195558e3130d69940 67e81a & -26.340 & 4.212\\
167 & 8 & 8 & ecad8e1a6be074ff88322c5b3cd5680dc82 5a1198 & -26.393 & 4.656\\
168 & 9 & 8 & 37f80dbe33864f68ef1fa5a951eeecd274d 5c6c506 & -25.421 & 4.324\\
169 & 9 & 9 & 18a745f218b371c6f21132f7f2f3ec9d290 f9eaae6f & -25.473 & 3.636\\
170 & 9 & 9 & 3835d25fe470f32ed7c4ccabe4f2b5e6601 1584a5bb & -25.524 & 4.234\\
171 & 9 & 9 & 20acaa4d24c6028139c5fd39f3065ca87cd 082f5f84 & -25.575 & 4.171\\
172 & 9 & 9 & 9c92f90c3ec2109c08862ec8ea5be45911d 7abb6143 & -25.626 & 3.928\\
173 & 9 & 9 & 1e58f6cad6917eeaee691536d57df81c5cb 901c43387 & -25.676 & 4.025\\
174 & 9 & 9 & 6c99808556a9e44f04a4af397f90dac63b5 c151f770 & -25.726 & 3.797\\
175 & 9 & 9 & 810552f57861b5543b90c9bc298de721699 f922627 & -25.776 & 3.923\\
176 & 9 & 9 & ac277413353446ebbec34fbda6a08305ea7 07e8b14a3 & -25.825 & 3.792\\
177 & 9 & 9 & bfd3bffc44db1369bde8c4956de06a2f3cc e38a9d0f9 & -25.875 & 4.366\\
178 & 9 & 9 & c40a317538cacb189615811a82f8a6da26c bc12fff85 & -25.924 & 3.905\\
179 & 9 & 9 & 755e7001560439f469090f9492191af2766 0ba19b2555 & -25.972 & 3.953\\
180 & 10 & 9 & 20e89f547a266727ad2c0e2dfbfab4eb790 0d6f11e714 & -26.021 & 4.147\\
181 & 9 & 9 & 112569db006c60067a7aee0fc75d29142a7 734da259170 & -26.069 & 4.143\\
182 & 10 & 9 & 55c099a8f91f8000d786cd73ce63b798a96 866a94bab6 & -26.117 & 4.434\\
183 & 10 & 9 & 567d0f51bc62247232345e7bd5c5073a4b4 d5002d822d & -26.164 & 3.946\\
184 & 10 & 9 & 9fb510fddd6402a513b7317c6506389a1e0 59a4b11bc65 & -26.212 & 4.121\\
185 & 10 & 9 & 992f9283fb8240a96fc5d5862d296463ce5 debb71d8ccd & -26.259 & 4.055\\
186 & 10 & 9 & 2efef5d4dde19fe9026e6db13acb718d287 83c9f8ef52a8 & -26.305 & 4.052\\
187 & 10 & 9 & 14f80f5c2591e69ce6e755251fbd683512c 2b6376eeedb & -26.352 & 4.899\\
188 & 10 & 9 & d22ffdd5f6a233a8bea58a16e81943e370e 6912d33c3136 & -26.398 & 3.936\\
189 & 10 & 9 & 123dbffccf13e5b1b781ed982dba92a278e 2573d64eaa9d & -26.444 & 4.663\\
190 & 10 & 9 & 2bfec663b8b80160e29f16b506d8b6e8955 261676066b042 & -26.490 & 4.246\\
191 & 9 & 9 & 11eac5b0b8ca5ad4c2d2744038c59fe6fe4 d07dd6c98b3f1 & -26.536 & 4.192\\
192 & 10 & 9 & ad3aaa94f48d92334e31e476fe2f033dffc 37f9042c32697 & -26.581 & 5.236\\
193 & 10 & 9 & aeb347c1d1da654e18f519cce85fc9df2c3 23bf65bfebc90 & -26.626 & 4.272\\
194 & 10 & 9 & 152b11e12881902387f696de45c5a36c92f 8a0ac77638caa7 & -25.756 & 4.200\\
195 & 10 & 9 & 1d47bac00fecaac330e5c6d93a68ce265e9 4ba9db0b030128 & -26.716 & 3.980\\
196 & 10 & 10 & e1be82e1e81af93cca3cd9dd75ec888046b 132b152c78404b & -25.845 & 4.436\\
197 & 10 & 10 & 1e71d8a9e4c75c8904f6dfea5e35495f00d ae91a0a1326ae05 & -25.889 & 3.716\\
198 & 10 & 10 & 2edba6c2298993d0ff35b502b939c8283fc 5bdf78ab63e79d4 & -25.933 & 3.959\\
199 & 10 & 10 & 213212bd5e84d6fbe8f059e2e39fbcb6399 b22ae39859b705f & -25.977 & 4.012\\
200 & 10 & 10 & 97f408aee9f17082e252ed9dd6354035128 4c780c85cf0cd0d & -26.021 & 4.016\\
201 & 10 & 10 & 1b8e271803e8f153e16ed49261efaeeda0d b5e9ac6ea62467f5 & -26.064 & 4.558\\
\bottomrule
\end{tabular}
\end{center}
\end{table}

\clearpage

\begin{table}[h]
\begin{center}
\caption{Results part III}
\label{tab:results3}
\tiny
\ttfamily
\rowcolors{2}{light-gray}{light-cyan}
\begin{tabular}{lccp{3.5cm}cc}
\toprule
$n$ & Old & New & Binary sequence in HEX & db & MF\\
\midrule
\showrowcolors
202 & 10 & 10 & 28896a25f8804e58cd76e40638bd0786ebc e96957888301b22a & -26.107 & 3.896\\
203 & 10 & 10 & 169c3c36a07652906d0deec88865527c4e8 c03e629baefe639e & -26.150 & 3.765\\
204 & 10 & 10 & cf67c809f660c8a7d9bc7aa4763d21c2105 135a2f235294545e & -26.193 & 4.130\\
205 & 10 & 10 & 159b65cba243039145c6500c7c65a7fe42d 0077ac87d1be36a54 & -26.235 & 4.511\\
206 & 10 & 10 & 12765377a7b55d926a14886701cfa80e3b0 5009f57a430e28cf8 & -26.277 & 4.546\\
207 & 10 & 10 & 3eda512837a306f55c4e618f6282b984c0a 22449efc32625e92f & -26.319 & 4.124\\
208 & 10 & 10 & d8c49d521383658069e764209165efb173a c434b843e15d4756b & -26.361 & 3.913\\
209 & 10 & 10 & 92bb7527a734817aab8268f1be66a10f871 3dc86dca35bd6dfe7 & -26.403 & 4.024\\
210 & 10 & 10 & 2e4b2cdb5d5d06708dddbda17e1097f8294 5cce2040c1e27438b7 & -26.444 & 4.416\\
211 & 10 & 10 & 3006f3f70992440f19518c5b08c22b12234 35582bfa5f3d26b7c0 & -26.486 & 3.853\\
212 & 10 & 10 & 2c1c395d9b2bad230839514a11bc85c866a 6389a27fac0fa2107e & -26.527 & 4.197\\
213 & 10 & 10 & 1ecb7a82ac839c1634e9a3c03160de3d009 43a2f549afed919bdc8 & -26.568 & 4.314\\
214 & 11 & 10 & 4391784839e2ba9e384fe40899ac6c696fd 5eba9949d3feb66914 & -26.608 & 4.004\\
215 & 11 & 10 & 54eba39307033259c5dd1ae000ba95a041b ef2b9be2d87f2e35ac2 & -26.649 & 4.294\\
216 & 11 & 10 & 70835039c47a166461a51e2e0bb2a4d756f 29f7f04bfbc9920127d & -26.689 & 4.160\\
217 & 11 & 10 & 11c2d59cc49c9469e9d6922094e8dba2617 501ec028d3fc705f3fcd & -26.729 & 3.999\\
218 & 11 & 10 & 12e61da78ed3e653f9cb64b6e8bf2145ee8 06877e7e76a8a819a9a1 & -26.769 & 4.162\\
219 & 11 & 10 & 2b37e41114e882ec5e59c25a9c57a203c0c 6b9699493c357c59ddf7 & -26.809 & 4.174\\
220 & 11 & 10 & 62a2bc0a38b1605f8321a7c8a13719d34a9 6f3446f6effc21148636 & -26.848 & 4.229\\
221 & 11 & 10 & f45bafe7673953bce07d5e74b7c041472ed a23e2cb7d49d32b1260b & -26.888 & 4.093\\
222 & 11 & 10 & d91d6ed119acea81c5f47ec6bd6d3be95a1 9ef9e465a0159070f764 & -26.927 & 4.046\\
223 & 10 & 10 & 4a70894496d298c01381155df82667e4cb3 21f97347c235e38170ae7 & -26.966 & 3.734\\
224 & 11 & 10 & 1e2e7c3249469a3537e2fe24612a5c9f520 5f4fa9a9bdec67bee2bb2 & -27.005 & 4.353\\
225 & 11 & 10 & 1dea3e715a9881e3e0054954159db182909 d36f961a4743e446b34ff1 & -27.044 & 4.609\\
226 & 11 & 10 & 32cc5e0c945afb4c12f3de9199312138c1d 88669015a8da3fd5474581 & -27.082 & 4.244\\
227 & 10 & 10 & 22ebf7574cc9779ebc090324b0cc61927b4 257f143313950f857ea553 & -27.121 & 4.251\\
228 & 11 & 10 & fa53a40f36c2f6374864b9c2c9ef7b2a284 c5fa79677ee1fea555b141 & -27.159 & 3.988\\
229 & 11 & 11 & a75ce55b5d23ecac9137d372bf947ea0c3a 221a1b30befb4b108fcf72 & -26.369 & 3.719\\
230 & 11 & 11 & f7332341300147a52cd1491971e815e65f1 036b8769a8aaf7159f2c47 & -26.407 & 3.854\\
231 & 11 & 11 & 2a808dc4d85ca8bd9682006611f9a363c8e 9ea6bebd2348d72c51a7c43 & -26.444 & 3.861\\
232 & 11 & 11 & 5656966e6e18f1dc48803edef7d24bb54ee d93e77334ebe02d3aab03d8 & -26.482 & 3.748\\
233 & 11 & 11 & 17d4bccaaf3086f2ab017b84178db7eec81 e279f5cbca7cbe68b5cd8da9 & -26.519 & 3.898\\
234 & 11 & 11 & 2007f0ac7762cac4e0e43831c4aa1a2240a 3dbb58536dead2cd534c61b6 & -26.556 & 4.009\\
235 & 11 & 11 & 479662251d2130781bcca255d6a87bbc42c 407c05258e8eac92838dbb66 & -26.594 & 3.830\\
236 & 11 & 11 & 240060c71fd710e97cbacb6a9de5b0aeb67 4353f352edc33609dd2f1337 & -26.630 & 3.901\\
237 & 11 & 11 & 19cc87e8436ee1b65ea0c8410034dd70a64 78e0d6a9d1575c5b89cb537 & -26.667 & 4.368\\
238 & 11 & 11 & 14cc4b3fad9b12199c1f4e96dfa8f5cd30e 7b50817c2f41ab8a362cf7a9a & -26.704 & 3.985\\
239 & 11 & 11 & 266ffb94a4f5bea647aa418dc69d151f1a2 9e6818ec9e5ee6e80f900720e & -26.740 & 3.560\\
240 & 12 & 11 & fe9c900f2c6ade00a1e0b104e12ce6b0fdd 2d54466a2146cfa2789ddb059 & -26.776 & 4.179\\
241 & 11 & 11 & 1c6b10f278e927d5b453595862437ec1f73 b713a9b86042153e2ec0054e8 & -26.812 & 4.170\\
242 & 11 & 11 & 83a8ab66dbf3e2e774631ee7e01f0d8957e 20e723dfc9512d2e3069a5eb4 & -26.848 & 4.425\\
243 & 11 & 11 & 4bb8a96e2929d4ed371fe8b99b623e16350 ffe48c167f6f3c22b9021952a8 & -26.884 & 4.251\\
244 & 11 & 11 & a0e8e4f0e137dc06decc6ad51bc2b11e12d 085843a610d47ffb4b20449b31 & -26.920 & 4.220\\
245 & 11 & 11 & 15e76533db7e903cd514700224afd24b2b4 033672fc1528f4308fa91ce0f1b & -26.955 & 3.877\\
246 & 11 & 11 & abfcd0b3f80b03974d8248b8a2b39b7fa5f 8c1ac676f61cd5fb7e729512bc & -26.991 & 3.756\\
247 & 12 & 11 & 55779587d0bc0a753acb17dbc71ae24d857 b967ef8529f3dfdd8466cb4c4cc & -27.026 & 3.966\\
248 & 11 & 11 & fee01b4a6b639587aefd2079b02a0fe1f51 a71ac419b55b5cb6666ebe7fa61 & -27.061 & 4.243\\
249 & 12 & 11 & e69021ef914b3dc4b720d828f4e78ad391e 8d0671766745d035a2ac6441440 & -27.096 & 4.444\\
250 & 12 & 11 & 1d82eb11b055fee7570494f67cadeeadebd 60e61c4e48a30f2b0495bd826c9e & -27.131 & 4.658\\
251 & 11 & 11 & 185b66591f9adfd4fcb9711a1ed865fd10e b1d31b5da95875bc4222eeef0b04 & -27.166 & 4.101\\
252 & 11 & 11 & 89e034220ae08d514bdaa363aaa4c2b7ed7 308c45bcdb44de44c3c7023cd85a & -27.200 & 4.033\\
253 & 12 & 11 & 1ed2db2821fbfae1870f40e99545e8e8f72 856cccdea1deb2ec37f91da769ac6 & -27.235 & 3.996\\
254 & 12 & 11 & 2e00e40a057f47b7764b2e91f2e1dc36752 0e74fc9857f5e9298cf5f6b6b1ac7 & -27.269 & 4.468\\
255 & 12 & 11 & 4e48792994ce3896f2363f70b53c43853aa aaed7c0b528101a17f4018136c933 & -27.303 & 3.993\\

\bottomrule
\end{tabular}
\end{center}
\end{table}

\begin{table}[h]
\begin{center}
\caption{Results part IV}
\label{tab:results4}
\tiny
\ttfamily
\rowcolors{2}{light-gray}{light-cyan}
\begin{tabular}{lccp{3.5cm}cc}
\toprule
$n$ & Old & New & Binary sequence in HEX & db & MF\\
\midrule
\showrowcolors
256 & 12 & 11 & 9b77e41cc0d9278fdd5a54b331946a53564 37b53baa902f780a61805078f2083 & -27.337 & 4.264\\
257 & 12 & 11 & 797b093ac095d0d53d4ce60de43928b1442 cb679e16ef7b80d5e76eddf8b45c9 & -27.371 & 4.116\\
258 & 12 & 11 & f19ccb67644aab3fac44bc02a8b7e62f7f4 ed5f6179f428da5d9b4983dc73c2c & -27.405 & 3.962\\
259 & 12 & 11 & 44c930912a770de24230e07dd434aca15a1 9580de8ab79ea8b37f1d90987cc182 & -27.438 & 4.534\\
260 & 12 & 11 & a50e2e9f7f7c415d2eb2cfab9be4ea46ab1 980f27c4cce6edc475ae09d216d382 & -27.472 & 4.370\\
261 & 11 & 11 & 10867d02c11be5517e4f4f5cbd4135e3f29 b15ffecae6e0d2d66479d064a678e79 & -27.505 & 3.920\\
262 & 12 & 11 & 128b8716cdead0448f3f6e7265ac6435c10 cefa2987f8c417035121484d40f452f & -27.538 & 4.198\\
263 & 12 & 11 & c5e44bf69b228002a7b29e90ef252a10727 0065cca0f0fea6d579de3acdc732ec & -27.571 & 4.430\\
264 & 12 & 11 & 838bacc30044321f263b7f2245bed79543d b437f5612e9d956a63389f8177469e7 & -27.604 & 4.264\\
265 & 12 & 11 & 1832e784ed916573709c6abcffb07ab5fea 0b3d5998abc9f0161ea37f7ad965a69 & -27.637 & 3.928\\
266 & 12 & 11 & 330bd8bc510f9d0045d9af80815954ee7d9 0a321066096387ec978dd496872d233c & -27.670 & 4.073\\
267 & 11 & 11 & df45ddba45d345ced1fb81f37be31a52a00 14bdf11cacc3a3e3589f5a5e490ca4f & -27.702 & 4.327\\
268 & 12 & 11 & 52b43d5792b8524e4edf6efb9b965597cf2 53c12f86ee5320c66efa122ff629c730 & -28.563 & 4.549\\
269 & 12 & 11 & 152b43d5792b8524e4edf6efb9b965597cf 253c12f86ee5320c66efa122ff629c730 & -27.767 & 4.388\\
270 & 12 & 11 & 21e5eea4f7cf140f85bea242277dde7bcd9 ca65dc4afae6f990be2a0678b4a966270 & -27.799 & 4.237\\
271 & 12 & 11 & 344052dfa92b00930cd10f1c58098a2a1cd afa3b9b962b0724c86837e291fc18ae56 & -27.832 & 3.838\\
272 & 12 & 11 & fcdde3a5833a16db6ed41cb0d2cc19c6fae cacb3a7ffa0dab51ba1cf6281b9d570fa & -27.864 & 4.393\\
273 & 12 & 12 & 44fd6cbb59dc119fee359596843d96f3db2 8c5eab59b0e2febc09f04560c206e4ab7 & -27.140 & 4.141\\
274 & 12 & 12 & 2084a897ae41a524bbff40ff05d12b96043 b5385d1fbb747137baa7399e5bc6c74dcd & -27.171 & 3.609\\
275 & 12 & 12 & 4092e0bb2873535a4739c7cf18ade8c273c 08cced32765fe95a0f45f6d66d564fafd5 & -27.203 & 3.980\\
276 & 12 & 12 & 3ad05cc5750b304c44d870be582126af4a6 7af40533e139a6afbdc6463ce0768206d8 & -27.235 & 3.563\\
277 & 12 & 12 & 183c883b5cb9366c4426e16eb70b50d862e ae61914bfa6805a78a29402e20758df57f & -27.266 & 3.912\\
278 & 12 & 12 & 3c474578ffdc1943abea11a0613a85b2970 d2665b3a7a4d4216113e233f348859d013c  & -27.297 & 4.077\\
279 & 12 & 12 & 7c815fd557ac5dce82804d1cf4b59b3ca8c e63cc72d2b270145a7220d82501fe8e049 & -27.328 & 3.983\\
280 & 12 & 12 & d27923a83fe74ff88a80248e14ad48d99ea 5ecf0d1f6d5dc6c18b773a8b167bb8c49ed  & -27.360 & 4.465\\
281 & 12 & 12 & 15a63b833b92922bda94f25432f9906e7d6 cb080b802c9d120101f66ae0d857078d3c6 3 & -27.391 & 3.823\\
282 & 12 & 12 & 4a0aa392e5296934d26cf8b8c007b8599be e514e4e7040326316ec4722f5abf06e4fef  & -27.421 & 3.949\\
283 & 12 & 12 & 131adc329deb49d4484ac1abfb560dd06c6 e9bb893abf288981e6107a08775c30a25f1 8 & -27.452 & 3.844\\
284 & 13 & 12 & 750721671bd43b577672bdbb85d72e9eb6d 8c2f778197470da082cefd9bfd061b61f63 6 & -27.483 & 3.961\\
285 & 12 & 12 & cfefe3c9a98b339b78784d6de752452df67 4bf76ef115867605ae316a075c142fe2451 a & -27.513 & 4.153\\
286 & 12 & 12 & 1e5f4df33a080874311aecb106e6bcf8aa4 e9fe29d34b36e7e427f23d71a8fbca3f4e2 d5 & -27.544 & 4.409\\
287 & 12 & 12 & 2d6fff0088403555d21c1be4513646065a4 2c2cde2742f397650ef9c8b432e8e5c0f6b 14 & -27.574 & 4.220\\
288 & 12 & 12 & d19696cc4945e90993653bcafae44afe6bf 3e1c872f1dfbf815e2a8c82f037d74dea9e 72 & -27.604 & 4.002\\
289 & 12 & 12 & 185bffdf5540ebb9cc935f8bcc4dabf1974 b54c1d5a9cf42b6383636e49c33ef889bc8 287 & -27.634 & 3.803\\
290 & 13 & 12 & 1954ea11d44ddb89d82999525ce70f41858 f24832d2eb011c1c81e6b0047edf5b71b47 e88 & -27.664 & 3.673\\
291 & 13 & 12 & 25e47f4fc982622a311acd9dcbbec0f2e9a 9f0cfda9dbe1823df5d7638c6d6194afe08 296 & -27.694 & 4.513\\
292 & 13 & 12 & a3b9c7d13899707f33a824ea27782dbb9bc e68925256b14fbb10795a0528e89010c3c7 68c & -27.724 & 4.072\\
293 & 13 & 12 & 17a57eb128c330309f040d6583d5e227709 75373e10547bf24a2e93d0cc8f996730a50 14c5 & -27.754 & 3.940\\
294 & 13 & 12 & 8dc9840017534c6eaf61cf1fbc58f568fa9 4403739476e14d72826e0bd289a4792fcd0 9c1 & -27.783 & 3.870\\
295 & 13 & 12 & 681faa2e4adff123767204b2af92e18b266 51a67e7c718c0619580c14a2b3f110cb42a 6da5 & -27.813 & 4.341\\
296 & 12 & 12 & fb1038796d64e801dc88d702cad89970a41 3091a431977d7be4ba8aa4eb721e3ba1409 acea & -27.842 & 4.238\\
297 & 12 & 12 & 6d0fb7212d379086a9e86c2a54fcc87ccfb a7ff6b9d4eca11b8f1e6c13eabafc448a39 a7d7 & -27.872 & 3.976\\
298 & 13 & 12 & 3aff50e3839b63e273c7ed3402274894e21 3b169fb37555558be5cc425a760b79f690c eba5f & -27.901 & 4.131\\
299 & 13 & 12 & 68ef75bbad75d36c63e30b296cd65f93e7e 0141f5bb84b81738c63ee47adab72a0f3b0 2cfb & -27.930 & 4.139\\
300 & 13 & 12 & b25be8354bc61f73a63b94ea06430063068 27e386dc8e36058b22aabb5a123b284c9fd f9504 & -27.959 & 4.365\\
\bottomrule
\end{tabular}
\end{center}
\end{table}

\clearpage

\bibliographystyle{IEEEtran}
\bibliography{refs}

\end{document}